\documentclass[aps,prd,twocolumn,showpacs,showkeys,amsmath,amssymb]{revtex4}
\usepackage{amsfonts}
\usepackage{amsmath}
\usepackage{graphicx}
\usepackage{subfigure}
\usepackage{dcolumn}
\usepackage{bm}
\usepackage{booktabs}
\usepackage[usenames,dvipsnames]{color}
\makeatletter
\newcommand{\figcaption}{\def\@captype{figure}\caption}
\newcommand{\tabcaption}{\def\@captype{table}\caption}

\newcommand{\Rmnum}[1]{\expandafter\@slowromancap\romannumeral #1@}
\def\hlinewd#1{%
  \noalign{\ifnum0=`}\fi\hrule \@height #1 \futurelet
   \reserved@a\@xhline}
\makeatother
\def\dab{\int^{\alpha_{max}}_{\alpha_{min}}d\alpha\int^{\beta_{max}}_{\beta_{min}}d\beta}
\def\qq{\langle\bar qq\rangle}
\def\GGa{\langle GG\rangle}
\def\GGb{\langle g_s^2GG\rangle}
\def\qGqa{\langle\bar  qGq\rangle}
\def\qGqb{\langle\bar qg_s\sigma\cdot Gq\rangle}

\def\f(s){\left[(\alpha+\beta)m_c^2-\alpha\beta s\right]}
\def\non{\\ \nonumber}

\usepackage{ulem}

\begin{document}

\title{$D^*\bar D^*$ molecule interpretation of $Z_c(4025)$}

\author{Wei Chen}
\email{wec053@mail.usask.ca}
\affiliation{Department of Physics
and Engineering Physics, University of Saskatchewan, Saskatoon, SK, S7N 5E2, Canada}
\author{T. G. Steele}
\email{tom.steele@usask.ca}
\affiliation{Department of Physics
and Engineering Physics, University of Saskatchewan, Saskatoon, SK, S7N 5E2, Canada}
\author{Meng-Lin Du} 
\email{duml@pku.edu.cn}
\author{Shi-Lin Zhu} 
\email{zhusl@pku.edu.cn}
\affiliation{Department of Physics and State Key Laboratory of Nuclear Physics and Technology,\\
Peking University, Beijing 100871, China }

\begin{abstract}
We have used QCD sum rules to study the newly observed charged state $Z_c(4025)$ as a hidden-charm $D^*\bar D^*$ molecular state with the 
quantum numbers $I^G(J^{P})=1^+(1^{+})$. Using a $D^*\bar D^*$ molecular interpolating current, we have calculated the two-point correlation 
function and the spectral density up to dimension eight at leading order in $\alpha_s$. The extracted mass is $m_X=(4.04\pm0.24)$ GeV. This result is compatible with the observed mass of $Z_c(4025)$ within the errors, which implies a possible molecule interpretation of this new resonance. We also predict the mass of the corresponding hidden-bottom $B^*\bar B^*$ molecular state: $m_{Z_b}=(9.98\pm0.21)$ GeV.
\end{abstract}



\maketitle


 \section{Introduction}\label{sec:intro}
After the observation of charged charmonium-like resonance $Z_c(3900)$~\cite{2013-Ablikim-p252001-252001}, the BESIII 
Collaboration recently discovered another charged structure $Z_c(4025)$ in the process $e^+e^-\to (D^*\bar D^*)^{\pm}\pi^{\mp}$ 
~\cite{2013-Ablikim-p-}. This new resonance, which has a mass of $M=(4026.3\pm2.6\pm3.7)$ MeV, lies very 
close to the $D^*\bar D^*$ threshold. Its width is $\Gamma=(24.8\pm5.6\pm5.7)$ MeV~\cite{2013-Ablikim-p-}. To date, the 
experiment has not determined the quantum numbers of the $Z_c(4025)$ resonance. Since it was observed in both the 
$D^*\bar D^*$ and the $h_c\pi$ channels, the quantum numbers of the charged $Z_c(4025)$ was argued to be 
$I^G(J^P)=1^+(1^+)$ while its neutral partner carries negative C-parity~\cite{2013-He-p-}. 

Similar to the other charged charmonium-like states $Z^+(4050)$, $Z^+(4250)$~\cite{2008-Mizuk-p72004-72004}, $Z^+(4430)$~\cite{2008-Choi-p142001-142001} 
and $Z_c(3900)$~\cite{2013-Ablikim-p252001-252001}, $Z_c(4025)$ cannot be a conventional $c\bar c$ state due to the charge it carries.
Molecular and  tetraquark configurations have recently been used to explore its underlying structure \cite{2013-He-p-}. 
In Ref.~\cite{2013-He-p-}, the authors have studied the mass spectrum of $Z_c(4025)$ and its pionic and radiative decays as a $D^*\bar D^*$ 
molecular state using the one-boson-exchange (OBE) model. They have also studied the $Y(4260)\to(D^*\bar D^*)^-\pi^+$ decay through the initial-single-pion-emission mechanism in Ref.~\cite{2013-Wang-p-}. $Z_c(4025)$ was also studied as a $[cu][\bar c\bar d]$ tetraquark with the quantum numbers $J^P=2^+$ using QCD sum rules~\cite{2013-Qiao-p-}. In Ref.~\cite{2013-Cui-p-}, the $D^*\bar D^*$ $J^P=1^+$ molecular current with a derivative has been studied in QCD sum rules and the extracted mass coincides with $Z_c(4025)$. 

There also exist other theoretical predictions of this new charged structure before its observation by BESIII
~\cite{2011-Sun-p54002-54002,2011-Chen-p34032-34032,2011-Chen-p34010-34010}. Ref.\cite{2011-Chen-p34010-34010} studied the $I^G(J^{PC})=1^+(1^{+-})$ charmonium-like tetraquark states, and predicted masses near the $D^*\bar D^*$ threshold and the possible decay patterns including the open-charm modes $D \bar D^*$, $ D^* \bar D^*$ and other hidden-charm modes. Up to now, the BESIII Collaboration has 
not reported the $D\bar D^*$ decay mode of $Z_c(4025)$. Right now, it seems that the $D^* \bar D^*$ molecule interpretation is slightly more natural.

At the hadronic level, the molecular states are commonly assumed to be bound states of two hadrons formed by the exchange of the color-singlet mesons. 
This configuration is very different from that of the tetraquark states, which are generally bound by the QCD force at the quark--gluon level.
In this work, we study $Z_c(4025)$ as a $D^*\bar D^*$ molecular state using QCD sum rules approach~\cite{1979-Shifman-p385-447,1985-Reinders-p1-1,2000-Colangelo-p1495-1576}.

Within the framework of the QCD sum rule, all the procedures such as the operator product expansion, the calculation of the Wilson coefficient 
and the Borel transform are very similar for the molecular-type current and tetraquark-type current. In principle, if we exhaust all the possible molecular-type 
currents and all the possible tetraquark-type of currents, we can rigorously show that these two sets of interpolating currents are equivalent by using a 
Fierz rearrangement~\cite{2007-Chen-p94025-94025,2009-Jiao-p114034-114034}.

However, there exists an important difference between one single molecular-type current and one single 
tetraquark-type current. By Fierz rearrangement, every single tetraquark-type current can be expressed as a linear combination 
of several (sometimes up to five) independent molecular-type currents. We can decompose the tetraquark interpolating current into 
these explicit molecular-type operators.
In the single molecular-type QSR, the color flow of the correlation function is quite simple and forms two closed loops. 
In the tetraquark correlator, there exist additional contributions from the non-diagonal correlator besides the many diagonal 
correlators as in the molecular-type QSR. Now the color flow is complicated, which is the interference and transition between 
different molecular structures~\cite{2010-Nielsen-p41-83}. In this respect, one well-known example is the light scalar-isoscalar sigma meson. 
The tetraquark-type current (or their combination/mixing) leads to a better mass prediction than the simple pion-pion molecular current~\cite{2007-Chen-p94025-94025}.

The paper is organized as follows. In Sect.~\Rmnum{2}, we calculate the correlation function and spectral density using the 
$D^*\bar D^*$ molecule current. In Sect.~\Rmnum{3}, we perform a numerical analysis and extract the mass of $Z_c(4025)$. 
The last section is a brief summary.

\section{QCD SUM RULE AND SPECTRAL DENSITY}\label{sec:current}
The starting point of QCD sum rules is the two-point correlation function
\begin{eqnarray}
\Pi_{\mu\nu}(q^{2})= i\int
d^4xe^{iq\cdot x}\langle0|T[J_{\mu}(x)J_{\nu}^{\dag}(0)]|0\rangle, \label{equ:Pi}
\end{eqnarray}
where $J_{\mu}(x)$ is the $D^*\bar D^*$ molecular interpolating current with $I^G(J^{P})=1^+(1^{+})$ 
\begin{eqnarray}
J_{\mu}=(\bar q_a\gamma^{\alpha}c_a)(\bar{c}_b\sigma_{\alpha\mu}\gamma_5q_b)-(\bar{q}_a\sigma_{\alpha\mu}\gamma_5c_a)(\bar c_b\gamma^{\alpha}q_b), \label{current}
\end{eqnarray}
in which $a, b$ are color indices and $q$ denotes an up or down quark. In principle, the anti-symmetric tensor operator $\bar{q}_a\sigma_{\alpha\mu}\gamma_5c_a$ can couple to 
both $J^P=1^+$ ($\bar{q}_a\sigma_{0i}\gamma_5c_a$ components) and $J^P=1^-$ ($\bar{q}_a\sigma_{ij}\gamma_5c_a$ components) channels. However, we can pick out the $1^-$ piece 
by multiplication with  the vector operator $\bar c_b\gamma^{\alpha}q_b$ so that the molecular operator $(\bar{q}_a\sigma_{\alpha\mu}\gamma_5c_a)(\bar c_b\gamma^{\alpha}q_b)$ carries the quantum numbers $J^P=1^+$ after contracting the Lorentz index. The molecule current in Eq.\eqref{current} contains both the charged components with $(\bar uc)(\bar cd)$ and $(\bar dc)(\bar cu)$ pieces and the neutral component with $(\bar uc)(\bar cu)$ and $(\bar dc)(\bar cd)$ pieces. For the neutral component, it carries negative C-parity and the quantum numbers should be $I^G(J^{PC})=1^+(1^{+-})$. However, we do not differentiate between $u$ and $d$ quarks in our analysis, so the charged component and the neutral  component are the same in QCD sum rules due to isospin symmetry. 

The correlation function in Eq. \eqref{equ:Pi} can be written as two independent Lorentz structures since $J_{\mu}$ is not a conserved current:
\begin{eqnarray}
\Pi_{\mu\nu}(q^{2})=\left(\frac{q_{\mu}q_{\nu}}{q^2}-g_{\mu\nu}\right)\Pi_1(q^2)+\frac{q_{\mu}q_{\nu}}{q^2}\Pi_0(q^2),
\end{eqnarray}
in which the invariant functions $\Pi_1(q^2)$ and $\Pi_0(q^2)$ are related to the spin-1 and spin-0 mesons, respectively. We focus on 
$\Pi_1(q^2)$ to study the $1^+$ channel in this work.
 
The correlation function in Eq. \eqref{equ:Pi} can be obtained at both the hadron level and the quark--gluon level. 
To determine the correlation function at the hadron level, we use the dispersion relation
\begin{eqnarray}
\Pi(q^2)=(q^2)^N\int_{4m_c^2}^{\infty}\frac{\rho(s)}{s^N(s-q^2-i\epsilon)}ds+\sum_{n=0}^{N-1}b_n(q^2)^n, \label{dispersionrelation}
\end{eqnarray}
where $b_n$ is the unknown subtraction constant which can be removed by taking the Borel transform. 
The lower limit of integration is the square of the sum of the masses of all current quarks (omitting the light quark mass). 
$\rho(s)$ is the spectral function
\begin{eqnarray}
\nonumber
\rho(s)&\equiv&\sum_n\delta(s-m_n^2)\langle0|J_{\mu}|n\rangle\langle n|J_{\nu}^{\dagger}|0\rangle
\\&=&f_X^2\delta(s-m_X^2)+ \mbox{continuum},  \label{Phenrho}
\end{eqnarray}
Here we adopt the pole plus continuum parametrization of the hadronic spectral density. The intermediate states $|n\rangle$ must have 
the same quantum numbers as the interpolating currents $J_{\mu}$. $|X\rangle$ is the lowest lying resonance with mass $m_X$ and it 
couples to the current $J_{\mu}$ via the coupling parameter $f_X$
\begin{eqnarray}
\langle0|J_{\mu}|X\rangle=f_X\epsilon_{\mu}, \label{coupling parameter}
\end{eqnarray}
where $\epsilon_{\mu}$ is the polarization vector ($\epsilon\cdot q=0$).

At the quark--gluon level, the correlation function can be calculated in terms of quark and gluon fields via the operator product 
expansion (OPE) method. We evaluate the correlation function up to dimension-eight condensate contributions at leading 
order in $\alpha_s$ using the same technique as in 
Refs.~\cite{2010-Chen-p105018-105018,2011-Chen-p34010-34010,2013-Du-p33104-33104,2013-Du-p14003-14003}. 
The spectral density is then obtained: $\rho(s)=\frac{1}{\pi}$Im$\Pi(q^2)$.

Sum rules for the hadron parameters are established by equating the correlation functions obtained at both the 
hadron level and quark--gluon level via quark--hadron duality. The Borel transform is applied to the correlation functions at 
both levels to remove the unknown constants in Eq.~\eqref{dispersionrelation}, suppress the continuum contribution, and improve 
the convergence of the OPE series. Using the spectral function defined in Eq. (\ref{Phenrho}), the sum rules can be written as
{\allowdisplaybreaks
\begin{eqnarray}
\nonumber f_X^2m_X^{2k}e^{-m_X^2/M_B^2}&=&\int_{4m_c^2}^{s_0}dse^{-s/M_B^2}\rho(s)s^k
\\&=&\mathcal{L}_{k}\left(s_0, M_B^2\right),
\label{sumrule}
\end{eqnarray}
where $s_0$ is the continuum threshold parameter and $M_B$ is the Borel mass. Then $m_X$ can be extracted by the ratio
\begin{eqnarray}
m_X=\sqrt{\frac{\mathcal{L}_{1}\left(s_0, M_B^2\right)}{\mathcal{L}_{0}\left(s_0, M_B^2\right)}}.
\label{mass}
\end{eqnarray}

In the following, we study the lowest lying hadron mass $m_X$ in Eq. \eqref{mass} as function of the continuum threshold 
$s_0$ and Borel mass $M_B$. We calculate the spectral density at the quark--gluon level including the perturbative term, 
quark condensate $\qq$, gluon condensate $\GGb$, quark--gluon mixed condensate $\qGqb$, four quark condensate $\qq^2$ and 
the dimension eight condensate $\qq\qGqb$:
\begin{eqnarray}
\nonumber
\rho(s)&=&\rho^{pert}(s)+\rho^{\qq}(s)+\rho^{\GGa}(s)+\rho^{\qq^2}(s)
\\ &&+\rho^{\qGqa}(s)+\rho^{\qq\qGqa}(s),
\end{eqnarray}
where
\begin{eqnarray}
\nonumber
\rho^{pert}(s)&=&\dab\f(s)^3
\non&&(1-\alpha-\beta)\Bigg\{\frac{m_c^2(\alpha+\beta-1)(5+\alpha+\beta)}{512\pi^6\alpha^3\beta^3}
\non&&+\frac{9(1+\alpha+\beta)\f(s)}{2048\pi^6\alpha^3\beta^3}\Bigg\},
\non
\rho^{\qq}(s)&=&-\frac{9m_c\qq}{64\pi^4}\dab(1-\alpha-\beta)
\non&&\frac{\f(s)\left[3m_c^2(\alpha+\beta)-7\alpha\beta s\right]}{\alpha\beta^2},
\non
\rho^{\GGa}(s)&=&\frac{\GGb}{1024\pi^6}\dab (1-\alpha-\beta)\Bigg\{
\non&&\frac{\f(s)\left[m_c^2(3+\alpha+\beta)+2\alpha\beta s\right]}{\alpha^2\beta}
\non&&+m_c^2(1-\alpha-\beta)\bigg[\frac{3\left[m_c^2(\alpha+\beta)-2\alpha\beta s\right]}{\alpha^3}
\non&&-\frac{(5+\alpha+\beta)\left[m_c^2(4\alpha+3\beta)-3\alpha\beta s\right]}{6\alpha\beta^3}
\non&&-\frac{(5+\alpha+\beta)\left[m_c^2(3\alpha+4\beta)-3\alpha\beta s\right]}{6\alpha^3\beta}\bigg]\Bigg\},
\non
\rho^{\qGqa}(s)&=&\frac{m_c\qGqb}{64\pi^4}\dab
\non&&\Bigg\{\frac{(1-\alpha-\beta)\left[3m_c^2(\alpha+\beta)-4\alpha\beta s\right]}{\beta^2}+
\non&&\frac{(2+7\alpha-2\beta)\left[3m_c^2(\alpha+\beta)-5\alpha\beta s\right]}{2\alpha\beta}\Bigg\},
\\
\rho^{\qq^2}(s)&=&\frac{5(s+2m_c^2)\qq^2}{48\pi^2}\sqrt{1-4m_c^2/s}\, ,
\non
\rho^{\qq\qGqa}(s)&=&\frac{\qq\qGqb}{48\pi^2}\int_0^1d\alpha
\non&&\Bigg\{\frac{3m_c^4(3-\alpha)}{\alpha^2(1-\alpha)}\delta'\left[s-\frac{m_c^2}{\alpha(1-\alpha)}\right]+
\non&&\frac{m_c^2(3\alpha^3-4\alpha^2-3\alpha+6)}{\alpha(1-\alpha)^2}\delta\left[s-\frac{m_c^2}{\alpha(1-\alpha)}\right]
\non&&+(3+2\alpha)H\left[s-\frac{m_c^2}{\alpha(1-\alpha)}\right]\Bigg\}. \label{spectral density}
\end{eqnarray}
}
in which $\alpha_{min}=\frac{1-\sqrt{1-4m_c^2/s}}{2}$, $\alpha_{max}=\frac{1+\sqrt{1-4m_c^2/s}}{2}$, 
$\beta_{min}=\frac{\alpha m_c^2}{\alpha s-m_c^2}$, $\beta_{max}=1-\alpha$, $m_c$ is the charm quark mass, and $H(\alpha)$ is the Heaviside step function. As is evident from the above expressions, our calculations are of leading order in $\alpha_s$.
Both the quark condensate $\qq$ and the quark--gluon mixed condensate $\qGqb$ are proportional to the charm quark mass $m_c$. They give important power 
corrections to the correlation functions. We ignore the chirally suppressed terms proportional to the light quark mass. Based on 
Ref. \cite{2010-Chen-p105018-105018} the contribution of the three gluon condensate $g_s^3\langle fGGG\rangle$ expected to be numerically 
small and has not been included in this work. The dimension-eight condensate $\qq\qGqb$ contains the delta function $\delta\left[s-\frac{m_c^2}{\alpha(1-\alpha)}\right]$ and its derivative. These terms compensate for the singular behavior of the spectral densities at the $s=4m_c^2$ threshold. 

\section{Numerical Analysis}\label{sec:NA}
The following QCD parameters are used in our 
analysis~\cite{2012-Beringer-p10001-10001,2001-Eidemuller-p203-210, 1999-Jamin-p300-303,2002-Jamin-p237-243,2011-Khodjamirian-p94031-94031}:
\begin{eqnarray}
\nonumber
&&m_c(m_c)=(1.23\pm0.09)\text{ GeV} \, , \non
&&m_b(m_b)=(4.20\pm0.07)\text{ GeV} \, , \non
&&\qq=-(0.23\pm0.03)^3\text{ GeV}^3 \, , 
\\ &&\qGqb=-M_0^2\qq\, ,
\non &&M_0^2=(0.8\pm0.2)\text{ GeV}^2 \, ,
\non&&\GGb=(0.88\pm0.14)\text{
GeV}^4 \, , \label{parameters}
\end{eqnarray}
where the charm and bottom quark masses are the running mass in the $\overline{MS}$ scheme. As mentioned earlier, we set the light quark 
masses $m_q=0$ in the analysis. The convention for the mixed condensate is consistent with Refs.
~\cite{2010-Chen-p105018-105018,2011-Chen-p34010-34010,2013-Du-p33104-33104,2013-Du-p14003-14003}, which have a sign difference from some other QCD sum rule studies because of the definition of the coupling constant $g_s$.

We define the pole contribution (PC) using the sum rules established in Eq. \eqref{sumrule},
\begin{eqnarray}
\text{PC}(s_0, M_B^2)=\frac{\mathcal{L}_{0}\left(s_0, M_B^2\right)}{\mathcal{L}_{0}\left(\infty, M_B^2\right)}, \label{PC}
\end{eqnarray}
which is the function of the continuum threshold $s_0$ and the Borel mass $M_B$. PC represents the lowest lying resonance contribution to the correlation 
function, which also includes the continuum and higher state contributions with $s_0\to\infty$. 

We begin with the analysis by determining the Borel window. A good mass sum rule requires a suitable working region of the Borel scale $M_B$. 
To obtain the lower bound on $M_B^2$, we let $s_0\to\infty$ and then study the OPE convergence in Fig. \ref{figOPE}. One notes that the quark condensate 
$\qq$ contribution is much bigger than other condensates and is therefore the dominant power correction. Besides the quark condensate, the quark--gluon mixed condensate 
$\qGqb$ also gives a significant contribution to the correlation function. From the expression for the spectral density in Eq. \eqref{spectral density}, the 
quark condensate and quark--gluon mixed condensate are proportional to the charm quark mass. The gluon condensate $\GGb$, four quark condensate $\qq^2$ and 
dimension-eight condensate $\qq\qGqb$ are smaller. However, they also give important corrections to the correlation function and stabilize the mass sum rules.
Requiring the quark condensate contribution be less than one third of the perturbative term contribution, while the quark--gluon mixed condensate contribution 
be less than one third of the quark condensate contribution, we obtain the lower bound on the Borel window $M_{min}^2=4.3\,{\rm GeV^2}$.
One may notice from Fig. \ref{figOPE} that the power corrections are small enough in the parameter region $M_B^2\geq4.3\,{\rm GeV^2}$ so that the OPE convergence is very good.
\begin{center}
\includegraphics[scale=0.7]{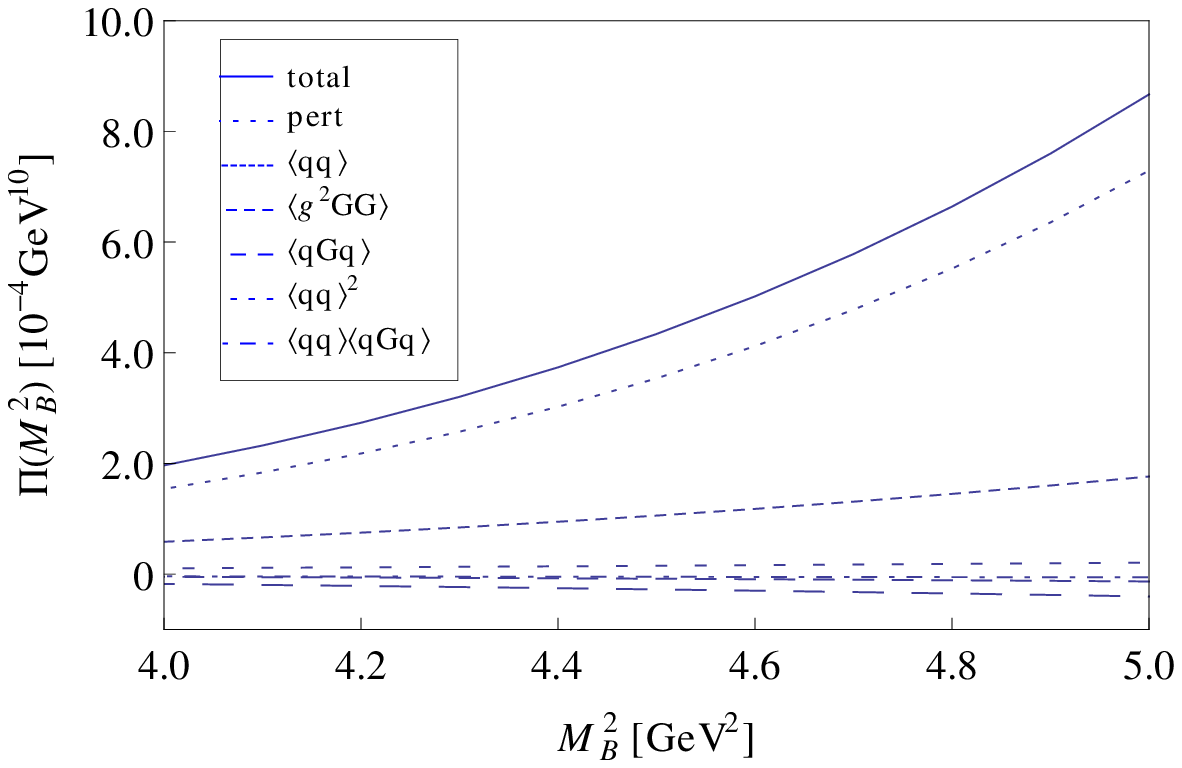}
\figcaption{The convergence of the OPE series with $s_0\to\infty$.} \label{figOPE}
\end{center}

\begin{center}
\includegraphics[scale=0.7]{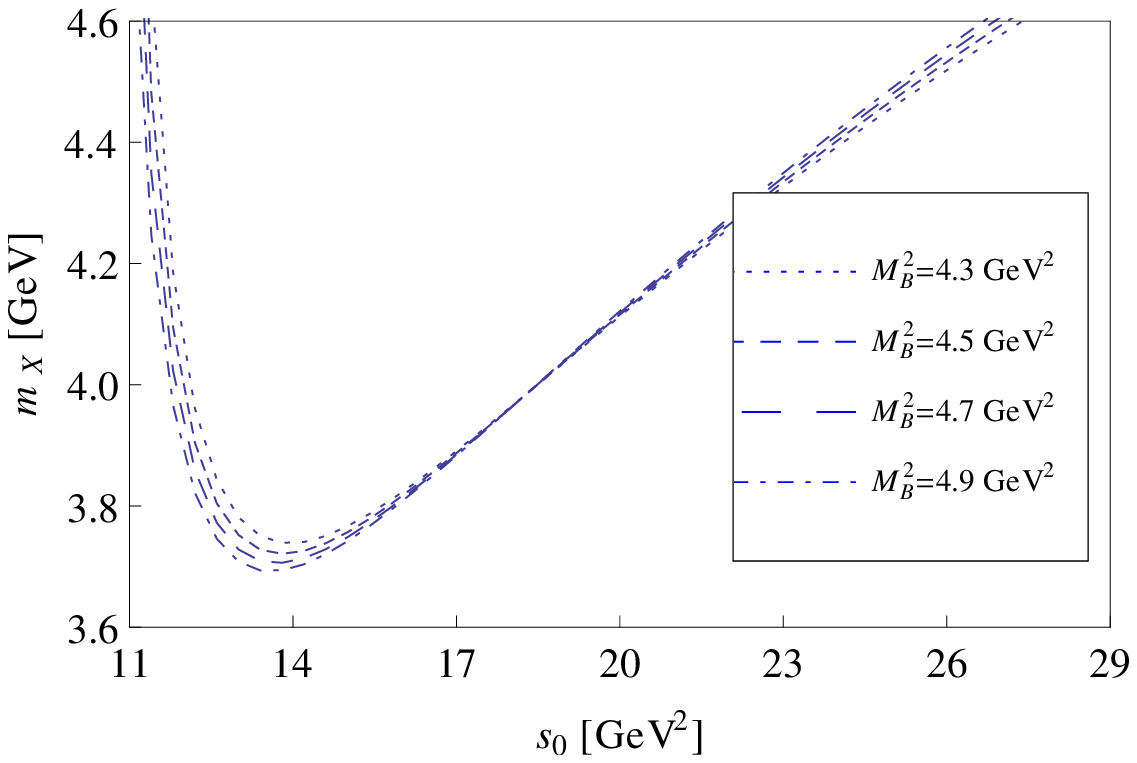}
\figcaption{The variation of $m_X$ with the continuum threshold $s_0$ in the Borel window $4.3\,{\rm GeV^2} \leq M_B^2\leq 4.9$ GeV$^2$.} \label{figms0}
\end{center}

The continuum threshold $s_0$ is also an important parameter in QCD sum rules. An optimized choice of $s_0$ is the value minimizing the 
variation of the extracted hadron mass $m_X$ with the Borel mass $M_B^2$. This is achieved by studying the variation of $m_X$ with $s_0$ 
in Fig. \ref{figms0} by varying the value of Borel mass from its lower bound $M_{min}^2$. One notes that these curves with a different value 
of $M_B^2$ intersect at $s_0=19$ GeV$^2$,  around which the variation of $m_X$ with $M_B^2$ is minimum. 
Then the upper bound on the Borel mass can be determined by studying the pole contribution defined in Eq. \eqref{PC}. 
We require that the pole contribution be larger than $10\%$, which results in the upper bound 
on the Borel mass $M_{max}^2=4.9\,{\rm GeV^2}$. We obtain the Borel window $4.3\,{\rm GeV^2}\leq M_B^2\leq4.9\,{\rm GeV^2}$ with the 
threshold value $s_0=19$ GeV$^2$.

Now we can perform the QCD sum rule analysis in the Borel window $4.3\,{\rm GeV^2} \leq M_B^2\leq 4.9$ GeV$^2$. In Fig. \ref{figmborel}, we show the 
variation of the extracted mass $m_X$ with the Borel mass $M^2_B$ using continuum thresholds $s_0=17$ GeV$^2$, $19$ GeV$^2$ and $21$ GeV$^2$ respectively. 
The mass curves are very stable in the Borel window around these threshold values. Finally, we extract the hadron mass:
\begin{eqnarray}
m_{X}=(4.04\pm0.24)~\text{GeV}, \label{massresult}
\end{eqnarray}
which is very well compatible with the mass of $Z_c(4025)$. This implies the possible $D^*\bar D^*$ molecule interpretation of this new resonance. 

Using this value of the hadron mass, we can also calculate the coupling parameter defined in Eq. \eqref{coupling parameter},
\begin{eqnarray}
 f_{X}=(0.012\pm0.005)~\text{GeV}^5. \label{fxresult}
\end{eqnarray}
This parameter represents the strength of the coupling of the current $J_{\mu}$ in Eq. \eqref{current} to the $Z_c(4025)$ resonance. 
The errors of our numerical results in Eq. \eqref{massresult} and Eq. \eqref{fxresult} involve the uncertainties in the heavy quark masses and the values of 
the quark condensate, quark--gluon mixed condensate, and gluon condensate in Eq. \eqref{parameters}. Other possible error sources such as truncation of the OPE 
series, the uncertainty of the continuum threshold value $s_0$ and the variation of the Borel mass $M_B$ are not taken into account.

\begin{center}
\includegraphics[scale=0.7]{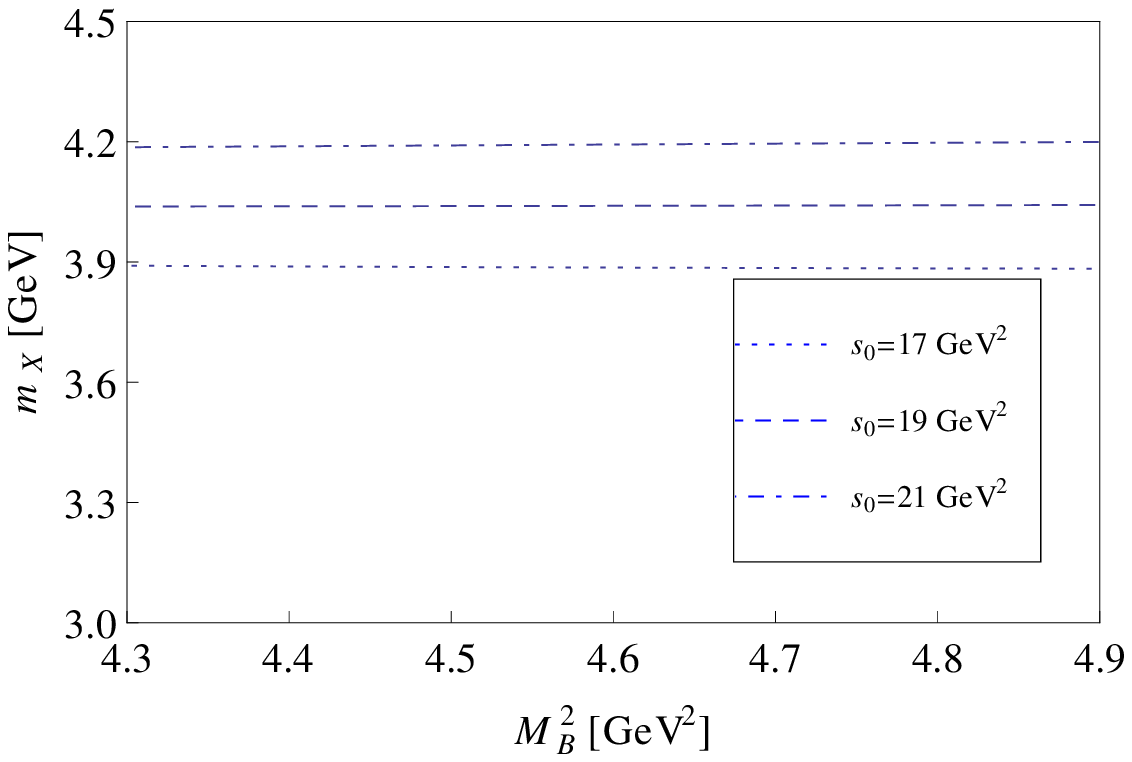}
\figcaption{The variation of $m_X$ with the Borel mass $M^2_B$ while $s_0=17$ GeV$^2$, $19$ GeV$^2$ and $21$ GeV$^2$.} \label{figmborel}
\end{center}

We can extend the analysis to the hidden-bottom $Z_b$ system, where $Z_b$ represents a $B^*\bar B^*$ molecular state with 
$I^G(J^{P})=1^+(1^{+})$. Using the same interpolating current in Eq. \eqref{current}, we repeat all the above analysis procedures 
with the replacement $m_c\to m_b$. To find a suitable working region of the Borel scale, we use the same criteria as in the $D^*\bar D^*$ 
system to study the OPE convergence and pole contribution. We find a Borel window 
$9.3\,{\rm GeV^2} \leq M_B^2\leq 11.6$ GeV$^2$ for the continuum threshold value $s_0=107$ GeV$^2$. 

\begin{center}
\includegraphics[scale=0.7]{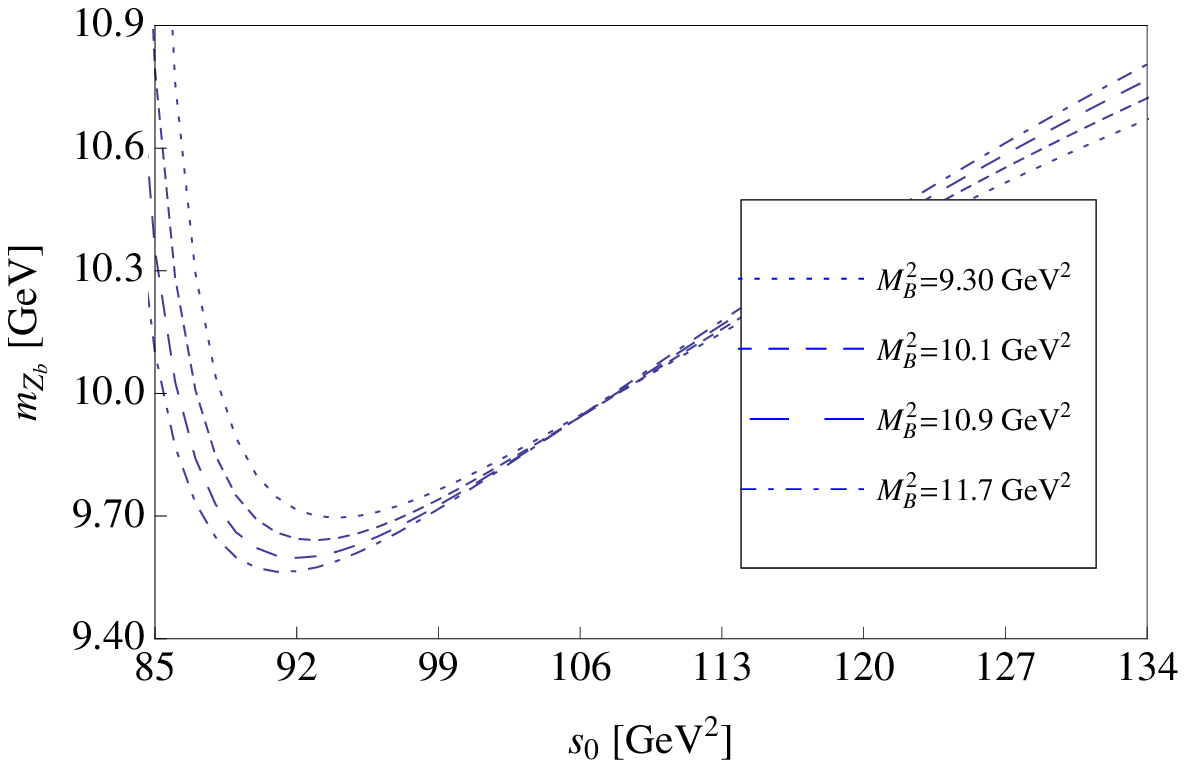}
\figcaption{The variation of $m_{Z_b}$ with the continuum threshold $s_0$ in the Borel window $9.3\,{\rm GeV^2} \leq M_B^2\leq 11.6$ GeV$^2$.} \label{figZbms0}
\end{center}
\begin{center}
\includegraphics[scale=0.7]{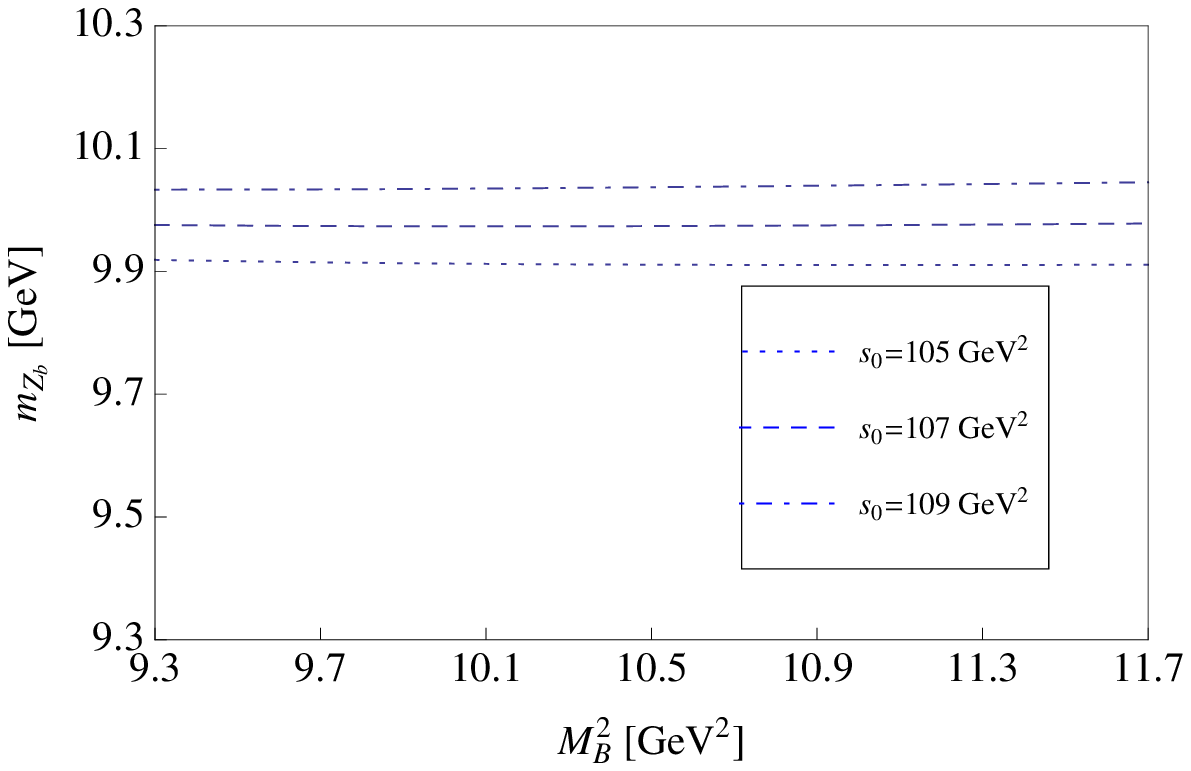}
\figcaption{The variation of $m_{Z_b}$ with the Borel mass $M^2_B$ while $s_0=105$ GeV$^2$, $107$ GeV$^2$ and $109$ GeV$^2$.} \label{figZbmborel}
\end{center}

We show the Borel curves of the extracted $Z_b$ mass with $s_0$ and $M_B^2$ in Figs. \ref{figZbms0} and \ref{figZbmborel}, 
respectively. In Fig. \ref{figZbms0}, the optimized value of the continuum threshold is chosen as $s_0=107$GeV$^2$, which 
minimize the variation of the $Z_b$ mass $m_{Z_b}$ with the Borel parameter $M_B^2$. This result is also shown in 
Fig.~\ref{figZbmborel}, in which the mass curve is very stable as a function of $M_B^2$ in the obtained Borel window.
Considering the same error sources as the $D^*\bar D^*$ system, we predict the mass and the coupling parameter 
of the $Z_b$ state to be
\begin{eqnarray}
m_{Z_b}=(9.98\pm0.21)~\text{GeV}, 
\\
f_{Z_b}=(0.003\pm0.001)~\text{GeV}^5.
\end{eqnarray}

\section{SUMMARY}\label{sec:SUMMARY}
The BESIII Collaboration has discovered $Z_c(4025)$ in the process $e^+e^-\to (D^*\bar D^*)^{\pm}\pi^{\mp}$ near the $D^*\bar D^*$
threshold. This new structure is a charged resonance and thus cannot be a conventional charmonium state. It is thus a candidate 
for an exotic hadron state. 

In Ref. \cite{2013-Cui-p-}, a $D^*\bar D^*$ molecular interpolating current with a derivative operator has been used to investigate 
the structure of $Z_c(4025)$ in QCD sum rules. In this paper, we use a different hidden-charm $D^*\bar D^*$ current with the 
quantum numbers $I^G(J^{P})=1^+(1^{+})$. We have calculated the correlation function and the spectral density up to dimension eight at 
leading order in $\alpha_s$, including the perturbative term, quark condensate $\qq$, quark--gluon mixed condensate $\qGqb$, gluon 
condensate $\GGb$, and the dimension-eight condensate $\qq\qGqb$ contributions. The quark condensate and the quark--gluon mixed condensate 
are proportional to the charm quark mass and are a larger contribution than the other condensates. The quark condensate 
is the dominant power correction to the correlation function. Other condensates are also important because they can improve the OPE convergence and stabilize the mass sum rules. 

After performing the QCD sum rule analysis, we extract the hadron mass $m_X=(4.04\pm0.24)$ GeV consistent with BESIII's result 
of the mass of the $Z_c(4025)$. 
Our result supports $Z_c(4025)$ resonance as an axial-vector $D^*\bar D^*$ molecular state. 
In principle, our result also contains the neutral partner of $Z_c(4025)$ with the quantum numbers $J^{PC}=1^{+-}$. However, it has the 
same mass with the charged state in QCD sum rules due to isospin symmetry. We have also studied the corresponding hidden-bottom 
$B^*\bar B^*$ molecular state and predicted the mass $m_{Z_b}=(9.98\pm0.21)$ GeV. Hopefully our investigation will be useful for the understanding 
of the structure of the newly observed charged state $Z_c(4025)$ and the future search of its neutral partner.

\section*{Acknowledgments}

This project was supported by the Natural Sciences and Engineering
Research Council of Canada (NSERC). S.L.Z. was supported by the
National Natural Science Foundation of China under Grants
11075004, 11021092, 11261130311 and Ministry of Science and
Technology of China (2009CB825200).


\begin{thebibliography}{22}
\expandafter\ifx\csname natexlab\endcsname\relax\def\natexlab#1{#1}\fi
\expandafter\ifx\csname bibnamefont\endcsname\relax
  \def\bibnamefont#1{#1}\fi
\expandafter\ifx\csname bibfnamefont\endcsname\relax
  \def\bibfnamefont#1{#1}\fi
\expandafter\ifx\csname citenamefont\endcsname\relax
  \def\citenamefont#1{#1}\fi
\expandafter\ifx\csname url\endcsname\relax
  \def\url#1{\texttt{#1}}\fi
\expandafter\ifx\csname urlprefix\endcsname\relax\def\urlprefix{URL }\fi
\providecommand{\bibinfo}[2]{#2}
\providecommand{\eprint}[2][]{\url{#2}}

\bibitem[{\citenamefont{Ablikim
  et~al.}(2013{\natexlab{a}})}]{2013-Ablikim-p252001-252001}
\bibinfo{author}{\bibfnamefont{M.}~\bibnamefont{Ablikim}} \bibnamefont{et~al.}
  (\bibinfo{collaboration}{BESIII collaboration}),
  \bibinfo{journal}{Phys.Rev.Lett.} \textbf{\bibinfo{volume}{110}},
  \bibinfo{pages}{252001} (\bibinfo{year}{2013}{\natexlab{a}}),
  \eprint{hep-ex/1303.5949}

\bibitem[{\citenamefont{Ablikim et~al.}(2013{\natexlab{b}})}]{2013-Ablikim-p-}
\bibinfo{author}{\bibfnamefont{M.}~\bibnamefont{Ablikim}} \bibnamefont{et~al.}
  (\bibinfo{collaboration}{BESIII collaboration})
  (\bibinfo{year}{2013}{\natexlab{b}}), \eprint{hep-ex/1308.2760}

\bibitem[{\citenamefont{He et~al.}(2013)\citenamefont{He, Liu, Sun, and
  Zhu}}]{2013-He-p-}
\bibinfo{author}{\bibfnamefont{J.}~\bibnamefont{He}},
  \bibinfo{author}{\bibfnamefont{X.}~\bibnamefont{Liu}},
  \bibinfo{author}{\bibfnamefont{Z.-F.} \bibnamefont{Sun}},
  \bibinfo{author}{\bibfnamefont{S.-L.} \bibnamefont{Zhu}}
  (\bibinfo{year}{2013}), \eprint{hep-ph/1308.2999}

\bibitem[{\citenamefont{Mizuk et~al.}(2008)}]{2008-Mizuk-p72004-72004}
\bibinfo{author}{\bibfnamefont{R.}~\bibnamefont{Mizuk}} \bibnamefont{et~al.}
  (\bibinfo{collaboration}{Belle}), \bibinfo{journal}{Phys. Rev.}
  \textbf{\bibinfo{volume}{D78}}, \bibinfo{pages}{072004}
  (\bibinfo{year}{2008}), \eprint{hep-ex/0806.4098}

\bibitem[{\citenamefont{Choi et~al.}(2008)}]{2008-Choi-p142001-142001}
\bibinfo{author}{\bibfnamefont{S.~K.} \bibnamefont{Choi}} \bibnamefont{et~al.}
  (\bibinfo{collaboration}{BELLE}), \bibinfo{journal}{Phys. Rev. Lett.}
  \textbf{\bibinfo{volume}{100}}, \bibinfo{pages}{142001}
  (\bibinfo{year}{2008}), \eprint{hep-ex/0708.1790}

\bibitem[{\citenamefont{Wang et~al.}(2013)\citenamefont{Wang, Sun, Chen, Liu,
  and Matsuki}}]{2013-Wang-p-}
\bibinfo{author}{\bibfnamefont{X.}~\bibnamefont{Wang}},
  \bibinfo{author}{\bibfnamefont{Y.}~\bibnamefont{Sun}},
  \bibinfo{author}{\bibfnamefont{D.-Y.} \bibnamefont{Chen}},
  \bibinfo{author}{\bibfnamefont{X.}~\bibnamefont{Liu}},
  \bibinfo{author}{\bibfnamefont{T.}~\bibnamefont{Matsuki}}
  (\bibinfo{year}{2013}), \eprint{hep-ph/1308.3158}

\bibitem[{\citenamefont{Qiao and Tang}(2013)}]{2013-Qiao-p-}
\bibinfo{author}{\bibfnamefont{C.-F.} \bibnamefont{Qiao}},
  \bibinfo{author}{\bibfnamefont{L.}~\bibnamefont{Tang}}
  (\bibinfo{year}{2013}), \eprint{hep-ph/1308.3439}

\bibitem[{\citenamefont{Cui et~al.}(2013)\citenamefont{Cui, Liu, and
  Huang}}]{2013-Cui-p-}
\bibinfo{author}{\bibfnamefont{C.-Y.} \bibnamefont{Cui}},
  \bibinfo{author}{\bibfnamefont{Y.-L.} \bibnamefont{Liu}},
  \bibinfo{author}{\bibfnamefont{M.-Q.} \bibnamefont{Huang}}
  (\bibinfo{year}{2013}), \eprint{hep-ph/1308.3625}

\bibitem[{\citenamefont{Sun et~al.}(2011)\citenamefont{Sun, He, Liu, Luo, and
  Zhu}}]{2011-Sun-p54002-54002}
\bibinfo{author}{\bibfnamefont{Z.-F.} \bibnamefont{Sun}},
  \bibinfo{author}{\bibfnamefont{J.}~\bibnamefont{He}},
  \bibinfo{author}{\bibfnamefont{X.}~\bibnamefont{Liu}},
  \bibinfo{author}{\bibfnamefont{Z.-G.} \bibnamefont{Luo}},
  \bibinfo{author}{\bibfnamefont{S.-L.} \bibnamefont{Zhu}},
  \bibinfo{journal}{Phys.Rev.} \textbf{\bibinfo{volume}{D84}},
  \bibinfo{pages}{054002} (\bibinfo{year}{2011}), \eprint{hep-ph/1106.2968}

\bibitem[{\citenamefont{Chen and Liu}(2011)}]{2011-Chen-p34032-34032}
\bibinfo{author}{\bibfnamefont{D.-Y.} \bibnamefont{Chen}},
  \bibinfo{author}{\bibfnamefont{X.}~\bibnamefont{Liu}},
  \bibinfo{journal}{Phys.Rev.} \textbf{\bibinfo{volume}{D84}},
  \bibinfo{pages}{034032} (\bibinfo{year}{2011}), \eprint{hep-ph/1106.5290}

\bibitem[{\citenamefont{Chen and Zhu}(2011)}]{2011-Chen-p34010-34010}
\bibinfo{author}{\bibfnamefont{W.}~\bibnamefont{Chen}},
  \bibinfo{author}{\bibfnamefont{S.-L.} \bibnamefont{Zhu}},
  \bibinfo{journal}{Phys. Rev.} \textbf{\bibinfo{volume}{D83}},
  \bibinfo{pages}{034010} (\bibinfo{year}{2011}), \eprint{hep-ph/1010.3397}

\bibitem[{\citenamefont{Shifman et~al.}(1979)\citenamefont{Shifman, Vainshtein,
  and Zakharov}}]{1979-Shifman-p385-447}
\bibinfo{author}{\bibfnamefont{M.~A.} \bibnamefont{Shifman}},
  \bibinfo{author}{\bibfnamefont{A.~I.} \bibnamefont{Vainshtein}},
  \bibinfo{author}{\bibfnamefont{V.~I.}
  \bibnamefont{Zakharov}}, \bibinfo{journal}{Nucl. Phys.}
  \textbf{\bibinfo{volume}{B147}}, \bibinfo{pages}{385} (\bibinfo{year}{1979})

\bibitem[{\citenamefont{Reinders et~al.}(1985)\citenamefont{Reinders,
  Rubinstein, and Yazaki}}]{1985-Reinders-p1-1}
\bibinfo{author}{\bibfnamefont{L.~J.} \bibnamefont{Reinders}},
  \bibinfo{author}{\bibfnamefont{H.}~\bibnamefont{Rubinstein}},
  \bibinfo{author}{\bibfnamefont{S.}~\bibnamefont{Yazaki}},
  \bibinfo{journal}{Phys. Rep.} \textbf{\bibinfo{volume}{127}},
  \bibinfo{pages}{1} (\bibinfo{year}{1985})

\bibitem[{\citenamefont{Colangelo}(2000)}]{2000-Colangelo-p1495-1576}
\bibinfo{author}{\bibfnamefont{K.~A.} \bibnamefont{Colangelo},
  \bibfnamefont{Pietro}}, \bibinfo{journal}{Frontier of Particle Physics}
  \textbf{\bibinfo{volume}{3}} (\bibinfo{year}{2000}), \eprint{hep-ph/0010175}

\bibitem[{\citenamefont{Chen et~al.}(2007)\citenamefont{Chen, Hosaka, and
  Zhu}}]{2007-Chen-p94025-94025}
\bibinfo{author}{\bibfnamefont{H.-X.} \bibnamefont{Chen}},
  \bibinfo{author}{\bibfnamefont{A.}~\bibnamefont{Hosaka}},
  \bibinfo{author}{\bibfnamefont{S.-L.} \bibnamefont{Zhu}},
  \bibinfo{journal}{Phys. Rev.} \textbf{\bibinfo{volume}{D76}},
  \bibinfo{pages}{094025} (\bibinfo{year}{2007}), \eprint{hep-ph/0707.4586}

\bibitem[{\citenamefont{Jiao et~al.}(2009)\citenamefont{Jiao, Chen, Chen, and
  Zhu}}]{2009-Jiao-p114034-114034}
\bibinfo{author}{\bibfnamefont{C.-K.} \bibnamefont{Jiao}},
  \bibinfo{author}{\bibfnamefont{W.}~\bibnamefont{Chen}},
  \bibinfo{author}{\bibfnamefont{H.-X.} \bibnamefont{Chen}},
  \bibinfo{author}{\bibfnamefont{S.-L.} \bibnamefont{Zhu}},
  \bibinfo{journal}{Phys. Rev.} \textbf{\bibinfo{volume}{D 79}},
  \bibinfo{pages}{114034} (\bibinfo{year}{2009})

\bibitem[{\citenamefont{Nielsen et~al.}(2010)\citenamefont{Nielsen, Navarra,
  and Lee}}]{2010-Nielsen-p41-83}
\bibinfo{author}{\bibfnamefont{M.}~\bibnamefont{Nielsen}},
  \bibinfo{author}{\bibfnamefont{F.~S.} \bibnamefont{Navarra}},
  \bibinfo{author}{\bibfnamefont{S.~H.} \bibnamefont{Lee}},
  \bibinfo{journal}{Phys. Rep.} \textbf{\bibinfo{volume}{497}},
  \bibinfo{pages}{41} (\bibinfo{year}{2010}), \eprint{hep-ph/0911.1958}

\bibitem[{\citenamefont{Chen and Zhu}(2010)}]{2010-Chen-p105018-105018}
\bibinfo{author}{\bibfnamefont{W.}~\bibnamefont{Chen}},
  \bibinfo{author}{\bibfnamefont{S.-L.} \bibnamefont{Zhu}},
  \bibinfo{journal}{Phys. Rev.} \textbf{\bibinfo{volume}{D81}},
  \bibinfo{pages}{105018} (\bibinfo{year}{2010}), \eprint{hep-ph/1003.3721}

\bibitem[{\citenamefont{Du et~al.}(2013{\natexlab{a}})\citenamefont{Du, Chen,
  Chen, and Zhu}}]{2013-Du-p33104-33104}
\bibinfo{author}{\bibfnamefont{M.-L.} \bibnamefont{Du}},
  \bibinfo{author}{\bibfnamefont{W.}~\bibnamefont{Chen}},
  \bibinfo{author}{\bibfnamefont{X.-L.} \bibnamefont{Chen}},
  \bibinfo{author}{\bibfnamefont{S.-L.} \bibnamefont{Zhu}},
  \bibinfo{journal}{Chin.Phys.} \textbf{\bibinfo{volume}{C37}},
  \bibinfo{pages}{033104} (\bibinfo{year}{2013}{\natexlab{a}}),
  \eprint{hep-ph/1203.5199}

\bibitem[{\citenamefont{Du et~al.}(2013{\natexlab{b}})\citenamefont{Du, Chen,
  Chen, and Zhu}}]{2013-Du-p14003-14003}
\bibinfo{author}{\bibfnamefont{M.-L.} \bibnamefont{Du}},
  \bibinfo{author}{\bibfnamefont{W.}~\bibnamefont{Chen}},
  \bibinfo{author}{\bibfnamefont{X.-L.} \bibnamefont{Chen}},
  \bibinfo{author}{\bibfnamefont{S.-L.} \bibnamefont{Zhu}},
  \bibinfo{journal}{Phys.Rev.} \textbf{\bibinfo{volume}{D87}},
  \bibinfo{pages}{014003} (\bibinfo{year}{2013}{\natexlab{b}}),
  \eprint{hep-ph/1209.5134}

\bibitem[{\citenamefont{Beringer et~al.}(2012)}]{2012-Beringer-p10001-10001}
\bibinfo{author}{\bibfnamefont{J.}~\bibnamefont{Beringer}} \bibnamefont{et~al.}
  (\bibinfo{collaboration}{Particle data group}), \bibinfo{journal}{Phys.Rev.}
  \textbf{\bibinfo{volume}{D86}}, \bibinfo{pages}{010001}
  (\bibinfo{year}{2012})

\bibitem[{\citenamefont{Eidemuller and Jamin}(2001)}]{2001-Eidemuller-p203-210}
\bibinfo{author}{\bibfnamefont{M.}~\bibnamefont{Eidemuller}},
  \bibinfo{author}{\bibfnamefont{M.}~\bibnamefont{Jamin}},
  \bibinfo{journal}{Phys. Lett.} \textbf{\bibinfo{volume}{B498}},
  \bibinfo{pages}{203} (\bibinfo{year}{2001}), \eprint{hep-ph/0010334}

\bibitem[{\citenamefont{Jamin and Pich}(1999)}]{1999-Jamin-p300-303}
\bibinfo{author}{\bibfnamefont{M.}~\bibnamefont{Jamin}},
  \bibinfo{author}{\bibfnamefont{A.}~\bibnamefont{Pich}},
  \bibinfo{journal}{Nucl. Phys. Proc. Suppl.} \textbf{\bibinfo{volume}{74}},
  \bibinfo{pages}{300} (\bibinfo{year}{1999}), \eprint{hep-ph/9810259}

\bibitem[{\citenamefont{Jamin et~al.}(2002)\citenamefont{Jamin, Oller, and
  Pich}}]{2002-Jamin-p237-243}
\bibinfo{author}{\bibfnamefont{M.}~\bibnamefont{Jamin}},
  \bibinfo{author}{\bibfnamefont{J.~A.} \bibnamefont{Oller}},
  \bibinfo{author}{\bibfnamefont{A.}~\bibnamefont{Pich}},
  \bibinfo{journal}{Eur. Phys. J.} \textbf{\bibinfo{volume}{C24}},
  \bibinfo{pages}{237} (\bibinfo{year}{2002}), \eprint{hep-ph/0110194}

\bibitem[{\citenamefont{Khodjamirian}(2011)}]{2011-Khodjamirian-p94031-94031}
\bibinfo{author}{\bibfnamefont{A.} \bibnamefont{Khodjamirian}},
\bibinfo{author}{\bibfnamefont{Th.} \bibnamefont{Mannel}},
\bibinfo{author}{\bibfnamefont{N.} \bibnamefont{Offen}},
  \bibinfo{author}{\bibfnamefont{Y.-M.}~\bibnamefont{Wang}},
  \bibinfo{journal}{Phys.Rev.} \textbf{\bibinfo{volume}{D83}},
  \bibinfo{pages}{094031} (\bibinfo{year}{2011})


\end{thebibliography}

\end{document}